\documentclass[aps,showpacs,pre,twocolumn, preprintnumbers,amsmath,amssymb]{revtex4}
\usepackage{graphicx} 
\usepackage{dcolumn} 
\usepackage{bm} 

\begin{document}
\title{Myelin figures: the buckling and flow of wet soap}
\author{Ling-Nan Zou} 
\email{zou@uchicago.edu}
\affiliation{The James Franck Institute and Department of Physics, The University of Chicago, Chicago, IL 60637}
\date{\today}
\begin{abstract}
Myelin figures are interfacial structures formed when certain surfactants swell in excess water. Here, I present data and model calculations suggesting myelin formation and growth is due to the fluid flow of surfactant, driven by the hydration gradient at the dry surfactant/water interface; a simple model based on this idea qualitatively reproduces the various myelin growth behaviors observed in different experiments. From a detailed experimental observation of how myelins develop from a planar precursor structure, I identify a mechanical instability that may underlie myelin formation. These results indicate the mixed mechanical character of the surfactant lamellar structure, where fluid and elastic properties coexist, is what enables the formation and growth of myelins.
\end{abstract}
\maketitle

\section{Introduction}

A bit of soap left in a wet dish turns into a gelatinous mass --- the sight may be unexceptional, but the microscopic structure of wet soap (hydrated surfactant) can be exceptionally diverse. The disparate interactions of the surfactant molecules with themselves and with water leads to the formation of aggregate structures, from spheres to cylinders (spherical and worm-like micelles) to extended membranes (bilayers); these can further organize into yet more complex structured phases \cite{NetoSalinas2005}. Here, temperature, pressure, and concentrations --- of surfactants, solvents, salts, etc. --- are the principle coordinates of a potentially very complex phase diagram.

Research into lyotropic (concentration-dependent) surfactant/water phases have largely focused on the equilibrium phase diagram. The kinetic processes through which the non-equilibrium initial state of pure surfactant and pure water arrives at equilibrium has been subjected to relatively few studies. Due to the strong concentration gradients, the interface between pure surfactant and pure water can consist of multiple domains of different phases, as well as metastable structures whose origins are not understood, evolving in spatially and temporally complex ways  \cite{BuchananArraultCates1998, BuchananEgelhaafCates2000, BuchananStarrsEgelhaafCates2000, BuchananEgelhaafCates2001}.  But even in surfactants whose phase diagram is fairly simple, the non-equilibrium kinetics can result in striking interfacial instabilities such as myelin figures.

First described in 1854 by Rudolf Virchow, who pioneered the use of microscopy in clinical pathology \cite{Virchow1854}, myelins are $\mu$m-diameter cylinders consisting of concentrically stacked bilayers, and are formed when some poorly-soluble surfactants swell in excess water \cite{SakuraiKawamura1984, Sakurai&c1985, SakuraiSuzukiSakurai1990, BuchananArraultCates1998, BuchananEgelhaafCates2000, Haran&c2002, DaveSurveManoharBellare2003, ZouNagel2006}. Their sometimes dramatic growth (reaching lengths $\sim 100\ \mu$m in a few seconds), together with their sinuous figure, give myelins a strangely life-like appearance.  In the 150 years since their discovery, the origin of myelins have remained mysterious. The equilibrium structure of myelin-forming surfactants consists of planar surfactant bilayers stacked one on top of another, separated by thin layers of water. Myelins, with tightly curled bilayers, have large bending elastic energies. Why then, do these seemingly energetically unfavorable structures form? Superficially, myelin formation, with the growth of finger-like structures from an initially flat interface, is reminiscent of non-equilibrium fingering processes such as the Mullins-Sekerka instability of a solidification front \cite{MullinsSekerka1962}. But a successful model for myelin formation based on such instabilities has yet to be introduced.

\begin{figure}[t]
\center
\includegraphics[width=2.5in]{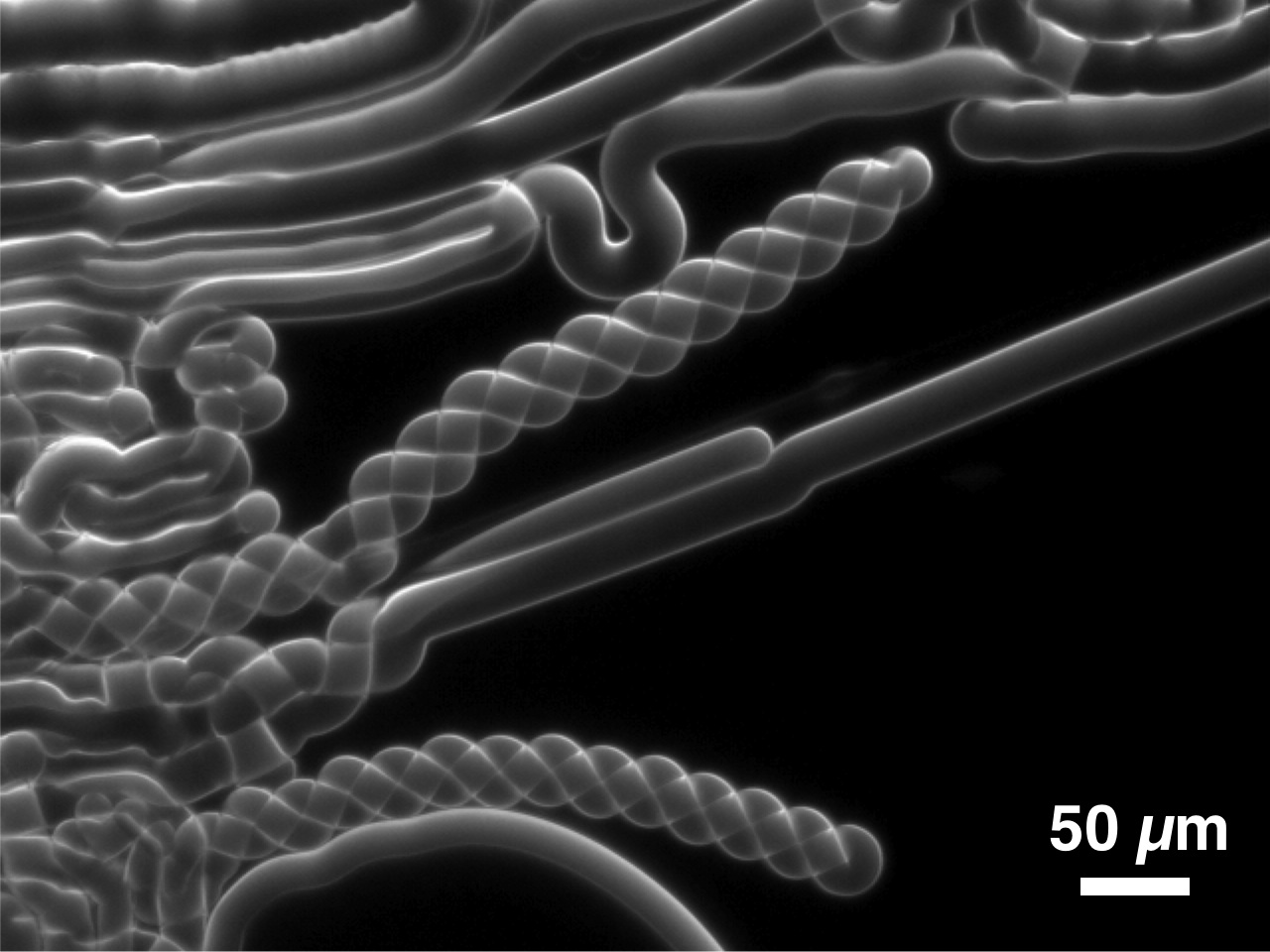}
\caption{Myelin figures viewed using dark field microscopy.}
\label{Myelin_Dark_Field}
\end{figure} 

The present paper proposes an alternative scenario. Due to the large concentration gradients at the concentrated surfactant/water interface, strong currents (of water and/or surfactant) must be present; these can focus large stresses at the interface, possibly leading to mechanical instabilities. I suggest myelin formation is due to such a mechanical instability. Interestingly, because the mechanical properties of the surfactant lamellar structure is anisotropic --- depending on the direction, it behaves either as a fluid or as an elastic solid, the myelin instability is one that has both fluid and elastic characteristics. 

This paper is organized as follows. Section II describes three distinct myelin growth experiments; their results are complementary, and together brings a fuller, more detailed picture of how myelins form and grow. In Section III, motivated by the varying myelin growth behaviors observed in these experiments, I introduce a simple model for myelin growth based on the \textit{fluid} flow of surfactant. The driving force behind this flow is the hydration gradient at the surfactant/water interface. Compared with experiment, this model qualitatively reproduces the observed myelin growth behaviors. Finally, Section IV describes in detail the formation of a myelin from a planar lamellar structure as observed in one of the experiments. I identity a mechanical instability which may underly myelin formation, and which gives good agreement with experimentally observed morphology and dynamics. Here the coexistence of \textit{fluid} and \textit{elastic} properties in the surfactant lamellar structure is crucial: in order for a myelin to form, the surfactant lamellar structure has to flow as a fluid and buckle as an elastic solid.

\section{Experimental Methods}

Some of the results from the experiments described below have been previously reported \cite{ZouNagel2006}; briefly summarized, these are: (1) myelins form laterally from the edges of their parent lamellar structures; (2) myelin growth is due to the flow of surfactant from the base and follow different time-dependences in different experiments; and (3) myelins grow and persist only in the presence of thermodynamic or mechanical stress. In Sections III and IV, I will revisit these results, present additional data, and compare experimental results with theoretical models. For this Section, I will restrict myself to describing the details of various experimental methods.

\begin{figure}[tb]
\center
\includegraphics[width=3in]{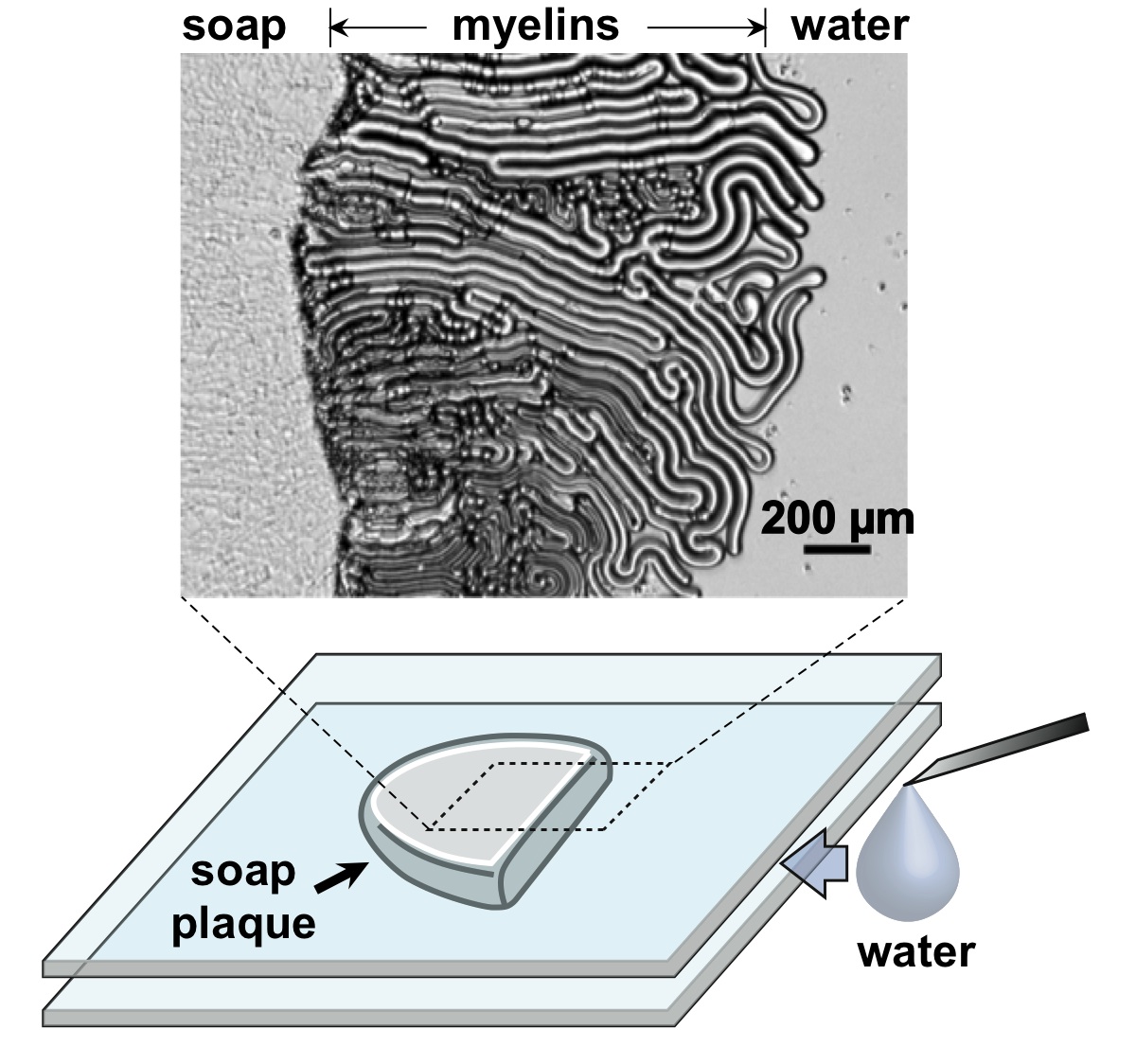}
\caption{(color online) The contact experiment. Surfactant is sandwiched between glass slides, and water (from a syringe) is introduced into the gap via capillary suction. Myelins form rapidly along the interface upon contact between water and surfactant.}
\label{contact_expt_setup}
\end{figure}

\subsection{Contact Experiment}

The simplest method to produce myelins to bring into contact a mass of appropriate surfactant with water. The surfactant I use is dilauroyl phospaditylcholine (DLPC), a common phospholipid. DLPC is dissolved in chloroform (concentration: $20 \ \textrm{mg/mL}$); a drop of the solution is cast onto a clean glass surface, and dried in dry nitrogen overnight. Just before use, the surfactant plaque is softened by brief exposure to the humidity from a warm water bath. The softened surfactant is carefully scrapped off and sandwiched between two glass slides, using a thin (0.15 mm) glass coverslip as spacer. Water is then introduced into the gap via capillary action, and contacts the surfactant plaque along an extended interface. Upon contact, myelins in dense bundles form and rapidly grow along the water/surfactant interface (Fig.~\ref{contact_expt_setup}).

To measure myelin growth, the experiment is observed via digital video-equipped dark-field microscopy in low magnification. The long thin myelins strongly scatter the indirect illumination, and appears bright while the bulk mass of surfactant remains dark. Thus the brightness of the video field is proportional to the amount of myelins formed. From the the average brightness of the myelin bundles, the average length of the myelins can be computed. This allows myelin growth to be monitored with a much higher temporal resolution than is practical by manual measurement.

\subsection{Immersion-and-Puncture Experiment}

\begin{figure}[tb]
\center
\includegraphics[width=3.25in]{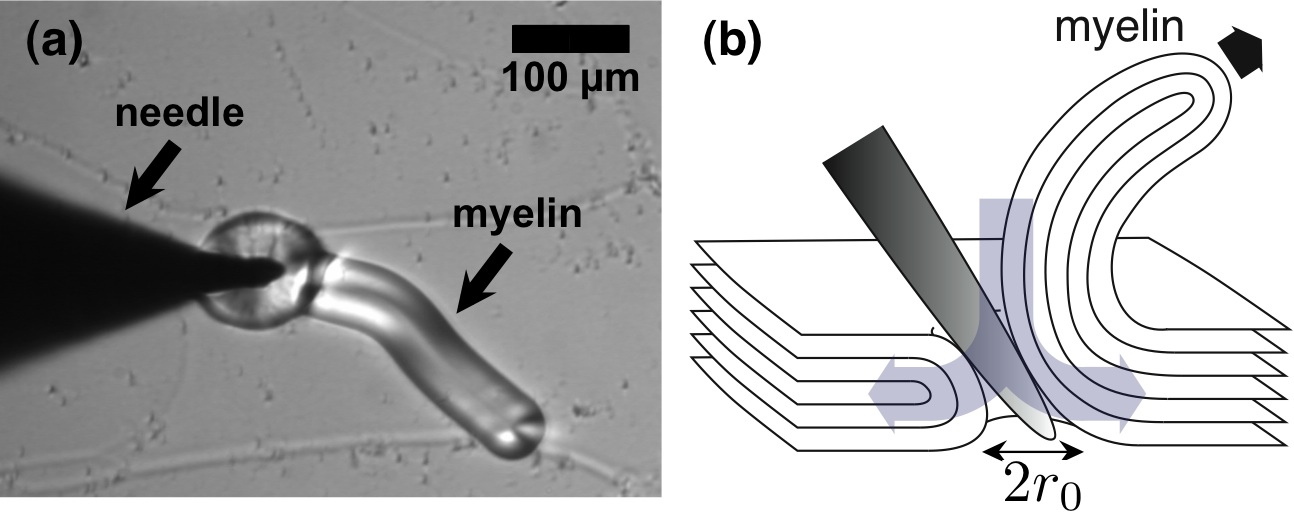}
\caption{(color online) (a) The immersion-and-puncture experiment as observed under the microscope. (b) Schematic illustration of the immersion-and-puncture experiment; presumably, the puncture as a hole of some effective size $r_0$ which eases water invasion into the surfactant plaque.}
\label{immersion_expt_setup}
\end{figure} 

Contact experiments as described previously do not in general have a good control over the structure of the concentrated surfactant plaque. Ideally, one would like to start with a well-ordered lamellar initial structure and observe if and how myelins develop.

To prepared the ordered initial structure, DLPC in chloroform is cast onto a clean glass-bottomed petri dish. Once the solvent has evaporated, the resultant surfactant plaques are incubated at $60^\circ \textrm{C}$ and 100\% relative humidity (r.h.) for up to two days. At the end of this process, the plaques anneal to form well-ordered planar structures with mm-size domains. The plaques are cooled and dehydrated to room temperature at 30\% r.h. The dish holding the surfactant plaques is then carefully flooded with water. Surprisingly, on immersion, few myelins are formed from the well-ordered surfactant plaques; this is in contrast to the dense thickets of myelins form on the surfactant/water interface in contact experiments.

However, myelins can be induced to form by puncturing the planar top surface of the surfactant plaques with a sharp needle (I use a tungsten or stainless steel electrophysiology electrode mounted on a micromanipulator). On pressing the needle into the plaque, one or more myelins will usually form at the site of the puncture and slowly grow in length. Presumably the puncture facilitates water invasion into the surfactant plaque  (Fig.~\ref{immersion_expt_setup}). As a clear illustration of the lateral fluidity of the surfactant lamellar structure, by moving the needle laterally across the plaque, the myelin can be readily dragged along without structural disruption. For earlier times, the growing myelin ``crawls'' along the surface, and its contour can be easily traced by a spline curve. As the myelin grows longer, its contour becomes convoluted and no longer quasi-2D; at this point the myelin's length can no longer be reliably measured.

\subsection{Drying Drop Experiment}

The drying drop experiment is another way to grow myelins in a controlled way from a well-ordered planar initial structure. The surfactant used here is dimyristol phospaditylcholine (DMPC); it is identical to DLPC except it has alkane tails groups that are two carbon atoms longer. Dry DMPC powder is dissolved in ethanol at a concentration of 50 mg/mL. A small quantity (10 $\mu$L) of the DMPC solution is pipetted into a microcentrifuge tube, then rapidly and throughly mixed with deionized water to form a vesicle suspension with final concentration of 0.5 mg DMPC/mL. This method consistently produces a fine suspension with a mean vesicle diameter of $\sim50$ nm as determined by dynamic light scattering. The small amount of ethanol in the suspension does not appear to influence myelin formation in these experiments.  The same phenomenon can be observed in experiments using ethanol-free vesicle suspensions produced (laboriously) by multiple extrusions across porous membranes.
 
\begin{figure}[tb]
\center
\includegraphics[width=2.5in]{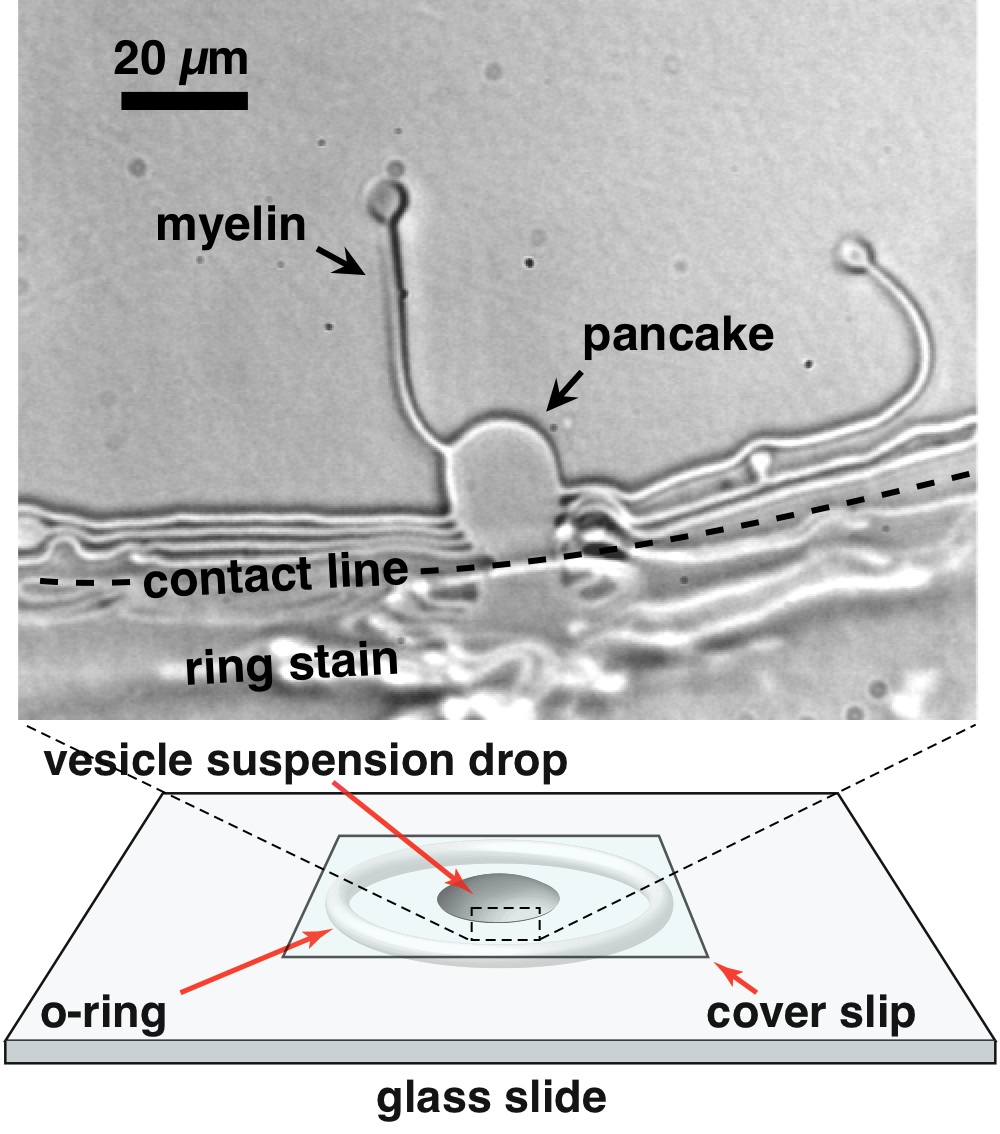}
\caption{(color online) Drying drop experiment: myelins grow from well-ordered lamellar pancakes located at the contact line of a slowly evaporating water drop.}
\label{drop_expt_setup}
\end{figure}

The experimental cell is constructed by epoxying a teflon o-ring on to a glass slide (Fig.~\ref{drop_expt_setup}).  A small drop of the DMPC suspension is placed inside the cell, and a glass cover slip is placed on top of the o-ring. The contact between the cover slip and the o-ring is intentionally left unsealed, allowing the drop to evaporate slowly: a 20 $\mu$L drop will take about 2 hours to dry completely. This geometry is best suited for observation using an inverted microscope; it can also be inverted, with the drop being pendant rather than recumbent, for observation using an upright microscope. The experiment is kept at $30^{\circ}$C.

The experiment works via the ``Coffee Stain Effect'' \cite{Deegan&c1997, Deegan&c2000}. As the drop dries, the suspended surfactant is deposited by capillary flow at the pinned contact line as a ring-stain.  The stain is mostly disordered, but small, well-ordered multilamellar ``pancakes'' are also formed right on the contact line. These pancakes grow slowly in size, and if one grows sufficiently large to invade back into the drop, it will form one or more myelins.

Drop evaporation can be halted midway through an experiment, by injecting a small drop of silicon oil near the o-ring/cover slip contact using a fine pipette. Capillary action will pull the oil into the o-ring/cover slip contact, forming an air-tight seal. Once the sealed cell equilibrates, in $\sim 1$ min, myelins stop growing and resorb into the pancakes from which they had developed earlier.

For this experiment to work effectively, it is important that the contact-line of the drying drop is pinned.  The surface characteristics of commercially available glass slides can vary considerbly, but the following treatment is found to consistently yield the desired properties. Glass slides are sonicated sequentially in toluene, methanol, and deionized water. They are then etched in a reactive ion etcher using $\textrm{CF}_4$, removing $\sim50$ nm of material, followed by a short etch with $\textrm{O}_2$. Finally they are ``aged'' in dry nitrogen for 10-12 hours before use.

Experiments are observed using transmitted bright field microscopy.  Real-time videos and still images are captured directly to computer using a CCD camera attached to the microscope.  High speed videos are captured using a Kodak Motion Corder camera and transferred to computer.

\section{Myelin growth: the fluid flow of soap}

Neglect for the moment, \textit{how} a myelin forms in the first place. In this Section, let us consider what causes a myelin, once formed, to grow. There is no reason \textit{a priori} to suppose the questions of how myelins form and how myelins grow can be separately considered; indeed, as I will suggest in Section III, myelin formation and myelin growth arise from the same phenomenon -- the fluid flow of surfactant driven by the gradient in water concentration. Nevertheless, it turns out to be both convenient and natural to consider these two questions separately.

\begin{figure}[t]
\center
\includegraphics[height=6.6in]{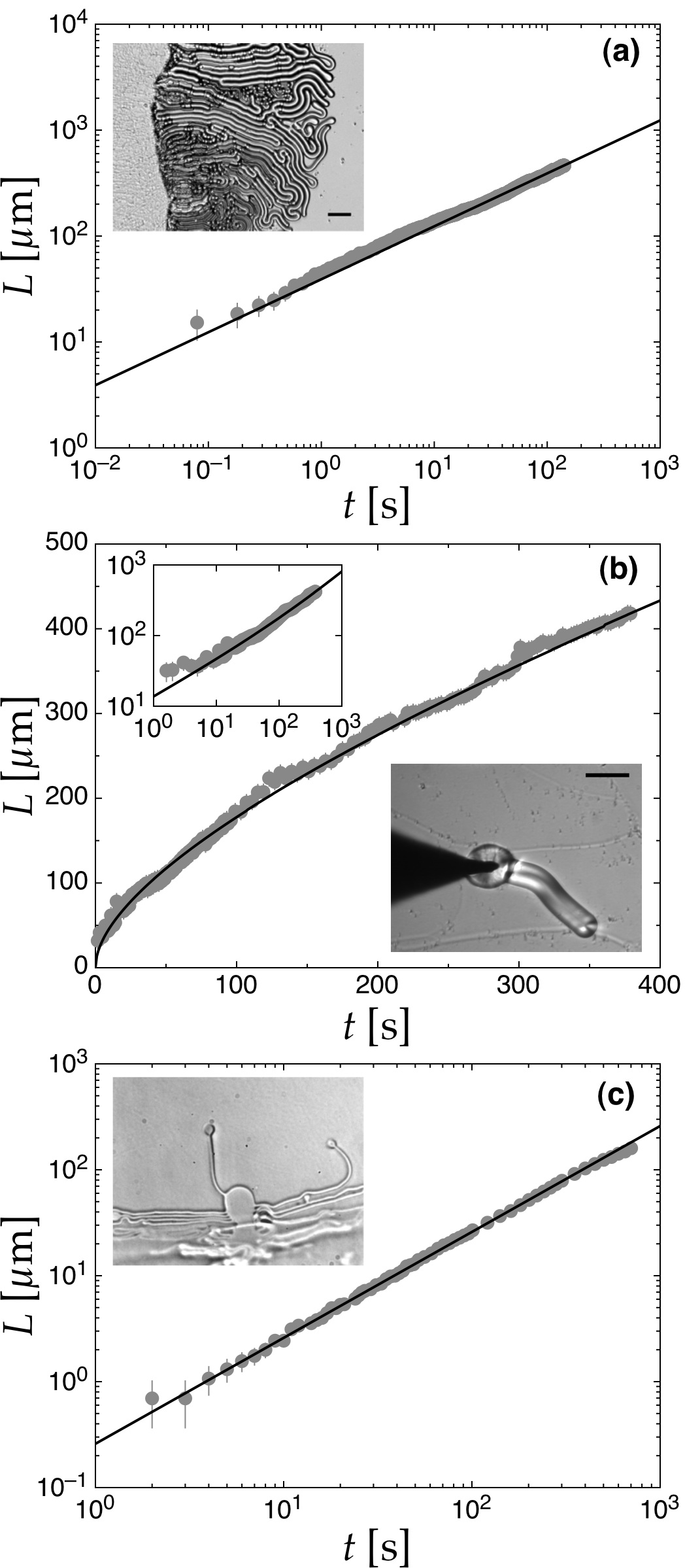}
\caption{Myelin growth has different behaviors in different experimetal geometries. (a) Contact experiments: solid line is a power law fit: $L \propto t^{1/2}$. (b) Immersion-and-punction experiments: solid line is a fit to Eq.~\ref{radial_L(t)}; insert shows the same data and fit plotted on logarithmic axes. (c) Drying drop experiments: solid line is a linear fit: $L \propto t$.}
\label{myelin_growth_L(t)}
\end{figure}

As previously reported, myelins grown in three experimental set-ups described in Section II follow different time dependences  \cite{ZouNagel2006}. In contact experiments, it has been well-documented that the length $L$ of the myelin bundles grows as $L \propto t^{1/2}$ (Fig.~\ref{myelin_growth_L(t)}a). By contrast, myelins in drying drop experiments grow linearly in time: $L \propto t$ (Fig.~\ref{myelin_growth_L(t)}c). Myelin growth in immerse-and-puncture experiments follow yet a third time-dependence which cannot easily described by any simple function (Fig.~\ref{myelin_growth_L(t)}b). A successful understanding of how myelins grow should describe all these different behaviors.

From the drying drop experiment, we know myelins grow by the flow of surfactant from the root; from the contact and immersion experiments,  we know applying osmotic stress can suppress myelin formation \cite{ZouNagel2006}. For myelin-forming surfactants, the surfactant chemical potential $\mu_\mathrm{s}$ decreases sharply with increasing $\Phi$, the local water fraction in the lamellar structure, reaching a minimum when $\Phi=\Phi_\mathrm{eq}\approx 0.5$, the fully swollen value \cite{ParsegianFullerRand1979}. If the surfactant sample is not uniformly hydrated, the hydration gradient $\nabla\mu_\mathrm{s}[\Phi(\mathbf{x})]$ is a thermodynamic pressure difference; here $\mu_\mathrm{s}$ is the surfactant chemical potential, and $\Phi(\mathbf{x})$ the local water fraction in the surfactant sample. \textit{Provided the surfactant molecules are mobile}, the hydration gradient can drive a surfactant flow from regions with low water content to regions of higher water content. In the simplest case, one can model the volumetric surfactant flow as
\begin{equation}
\mathbf{J}_\mathrm{s}=-\gamma_\mathrm{s} \nabla \mu_\mathrm{s}[\Phi(\mathbf{x})] \approx \gamma_\mathrm{s} \mu_\mathrm{s}^\mathrm{eq} \nabla \Phi(\mathbf{x}) \equiv \sigma \nabla \Phi(\mathbf{x}) \ ,
\label{flux_eq}
\end{equation}
where $\gamma_\mathrm{s}$ is the surfactant mobility and $\mu_\mathrm{s}^\mathrm{eq} \approx k_\mathrm{B} T$ is the fully-hydrated surfactant chemical potential; $\sigma \equiv \gamma_\mathrm{s} \mu_\mathrm{s}^\mathrm{eq}$ is therefore a diffusion constant. Myelins result when this flow breaks through the initial surfactant/water interface $S$, and the growth rate given by
\begin{equation}
\frac{dL}{dt} = \mathbf{J}_\mathrm{s}\cdot\hat{n}\big{\vert}_S = \sigma \nabla \Phi(\mathbf{x}) \cdot\hat{n}\big{\vert}_S \ ;
\label{growth_eq}
\end{equation}
here $L$ is the myelin length, and $\hat{n}$ is the outward unit normal on $S$.  Generally speaking, surfactant flow is not the only one present: water can flow too. The hydration profile $\Phi(\mathbf{x})$ can evolve in time, for example, as water permeates into the dehydrated surfactant,
\begin{equation}
\frac{\partial\Phi(\mathbf{x}, t)}{\partial t} = D \nabla^{2}\Phi(\mathbf{x}, t) \ ;
\label{diffusion_eq}
\end{equation}
here $D$ is the diffusion constant for water permeation. The full dynamics of myelin growth arises from the coupling of $\mathbf{J}_\mathrm{s}$ to $\Phi$. In particular, since transverse surfactant exchange between adjacent but unconnected bilayers is negligible in myelin-forming surfactants\cite{Israelachvili1992}, the surfactant diffusion constant $\sigma=(\sigma_\parallel, \sigma_\perp)$ is a tensor whose transverse component $\sigma_\perp$ is zero; therefore surfactant flow is possible only if the hydration gradient has a lateral component.  As I will show below, once the proper boundary and initial conditions are imposed,  this minimal set of ingredients are sufficient to reproduce qualitatively all the observed myelin growth behaviors.

It is useful here to have an estimate of the values of the transport coefficients $\sigma_{\parallel}$ and $D$. The lateral diffusion of surfactant molecules on bilayers has been well-studied. In the limit of small driving forces, one can identify $\sigma_\parallel$ with the bilayer's lateral surfactant diffusion constant: $\sigma_\parallel \approx 2-3 \  \mu\textrm{m}^2/\textrm{s}$. On the other hand, how water moves within a surfactant lamellar structure has not been studied in detail, and I rely instead on the following estimate: if the movement of water within a surfactant sample is rate-limited by the permeation of water across bilayers, then in the limit of low forcing, $D \approx Kd_\mathrm{w}$, where $K$ is the bilayer permeability and $d_\mathrm{w}$ is the inter-bilayer separation \cite{Tanner1978, DudkoBerezhkovskiiWeiss2004, CarruthersMelchior1983}.  For DMPC, $K \approx 2 \  \mu\textrm{m/s}$, $d_\mathrm{w}\approx 3\ \textrm{nm}$, yields $D \approx 6\times 10^{-3} \ \mu\textrm{m}^2/\textrm{s}$.

\subsection{Contact Experiment} 

In the contact experiment, myelins form at the extended interface between a large mass of dry surfactant and bulk water (Fig.~\ref{contact_expt_setup}). The situation may be simplified as a semi-infinite domain of surfactant ($x<0$) contacting a semi-infinite domain of water ($x>0$), with the interface at $x=0$.  The problem reduces to a system of two 1D equations,
\begin{equation}
\frac{\partial\Phi}{\partial t} = D\frac{\partial\Phi}{\partial x^{2}} \quad \textrm{and} \quad
\frac{dL}{dt} = \bar{\sigma}\frac{\partial\Phi}{\partial x}\bigg\vert_{x=0} \ ,
\end{equation}
governing water equilibration and myelin growth, respectively. Here, I have replaced $\sigma$ with a coarse-grained average $\bar{\sigma}$, since in these experiments, the surfactant sample is always very disordered, with small randomly oriented lamellar domains \cite{ZouNagel2006}. The initial and boundary conditions are
\begin{equation}
\Phi(x<0, 0) = \Phi(-\infty, t) = 0 \ , \quad \Phi(0, t) = \Phi_\mathrm{eq} \ .
\end{equation}
The solution for myelin growth $L(t)$ is given by
\begin{equation}
L(t) = \frac{2\bar{\sigma}\Phi_\mathrm{eq}}{\sqrt{\pi D}} \ t^{1/2},
\end{equation}
which has the same time-dependence as observed in experiment (Fig.~\ref{myelin_growth_L(t)}a).  The prefactor in front of $t^{1/2}$ is proportional to the equilibrium water volume fraction $\Phi_\mathrm{eq}$. Under the application of osmotic stress $\Pi$, $\Phi_\mathrm{eq}\rightarrow\Phi_\mathrm{eq}(\Pi)$ is suppressed; this implies myelin growth will also be suppressed, in qualitative agreement with experiments \cite{ZouNagel2006}. Using $\bar{\sigma} \approx \sigma_\parallel = 2\ \mu\textrm{m}^2/\textrm{s}$, $D = 6 \times 10^{-3}\ \mu\textrm{m}^2/\textrm{s}$, and $\Phi_\mathrm{eq}=0.5$ into the prefactor $m$, one obtains $15\ \mu\textrm{m}\cdot \textrm{s}^{-1/2}$ for the prefactor, which is consistent with experimentally measured values of $20-40 \ \mu\textrm{m}\cdot \textrm{s}^{-1/2}$. 

\subsection{Immersion Experiment}

In the immersion experiment, initially the hydration gradient is transverse to the bilayers of the well-ordered lamellar structure. However, the transverse surfactant diffusion constant is $\sigma_{\perp}=0$; without surfactant flow, myelins do not grow \cite{ZouNagel2006}. When the needle is pushed into the lamellar structure, the puncture acts as a point conduit for water invasion into the surfactant (Fig.~\ref{immersion_expt_setup}b). There, the hydration gradient will be radial-lateral in orientation, to which the lateral surfactant diffusion (controlled by $\sigma_{\parallel}$) can couple, leading to myelin growth.  One can approximate this situation as a 1D problem in polar coordinates.  The relaxation of the hydration profile and myelin growth are given by
\begin{equation}
\frac{\partial \Phi}{\partial t} = D \left[\frac{1}{r}\frac{\partial\Phi}{\partial r}+\frac{\partial\Phi}{\partial r^{2}}\right] \quad \textrm{and} \quad \frac{dL}{dt}=\sigma_{\parallel}\frac{\partial\Phi}{\partial r}\bigg\vert_{r=r_0} \ ,
\end{equation}
with the initial and boundary conditions
\begin{equation}
\Phi(r>r_0, 0) = \Phi(\infty, t) = 0 \ , \quad \Phi(r_0, t) = \Phi_\textrm{eq} \ ;
\end{equation}
here $2r_0$ is the effective size of the water conduit created by the needle. Solving for $L(t)$ yields \cite{Crank1980}
\begin{equation}\label{radial_L(t)}
L(t) = \frac{4\Phi_\mathrm{eq}\sigma_\parallel r_0}{\pi^2 D}\!\int_0^\infty \!\frac{1-e^{-u^2 D t/r_0^2}}{u^3\left[J_0^2(u)+Y_0^2(u)\right]} \ du \ .
\end{equation}\label{immersion_expt_L(t)}

Compared to data, Eq.~\ref{radial_L(t)} gives a reasonable fit (Fig.~\ref{myelin_growth_L(t)}b). Again, the solution contains a prefactor proportional to $\Phi_\mathrm{eq}$, indicating that myelin will be suppressed by applied osmotic stress. From the fit to Eq.~\ref{radial_L(t)}, one obtains $D/r_0^{2}\approx 0.01\ \textrm{s}^-1$ and $\sigma_{\parallel}r_0/D\approx 200\ \mu\textrm{m}$. If we take $\sigma_{\parallel}=2 \ \mu\textrm{m}^{2}/\textrm{s}$, then the fit implies $r_0\approx 1\ \mu\textrm{m}$ as the effective size of the puncture, a small but physically plausible value. The fit also implies $D\approx 10^{-2}\ \mu\textrm{m}^{2}/\textrm{s}$ as the diffusion constant for water permeation; this is consistent with the independent estimate of $D$ based on bilayer permeability.

\subsection{Drying Drop Experiment}  

In the drying drop experiment, myelins are formed from pancakes situated at the pinned contact line of a slowly evaporating drop (Fig.~\ref{drop_expt_setup}). Because the contact line is stationary, water influx across the contact line into the pancake must be balanced by water loss due to evaporation. The hydration profile $\Phi(\mathbf{x})$ is therefore static, and $dL/dt \propto \nabla\Phi\vert_S$ is a constant.  Myelin growth should be linear in time,
\begin{equation}\label{linear_growth}
L(t) =  \sigma_\parallel \nabla\Phi\vert_S \ t \ ,
\end{equation}
in agreement with experiment (Fig.~\ref{myelin_growth_L(t)}c). Fitting the experimentally observed $L(t)$ yields myelin growth rates of $0.2-0.5\ \mu\textrm{m}/\textrm{s}$. This implies at the strength of the hydration gradient $\nabla\Phi\vert_{S}$ is approximately $0.1-0.3\ \mu\textrm{m}^{-1}$ at the contact line, which is physically plausible.

\begin{figure*}[tb]
\center
\includegraphics[height=2.2in]{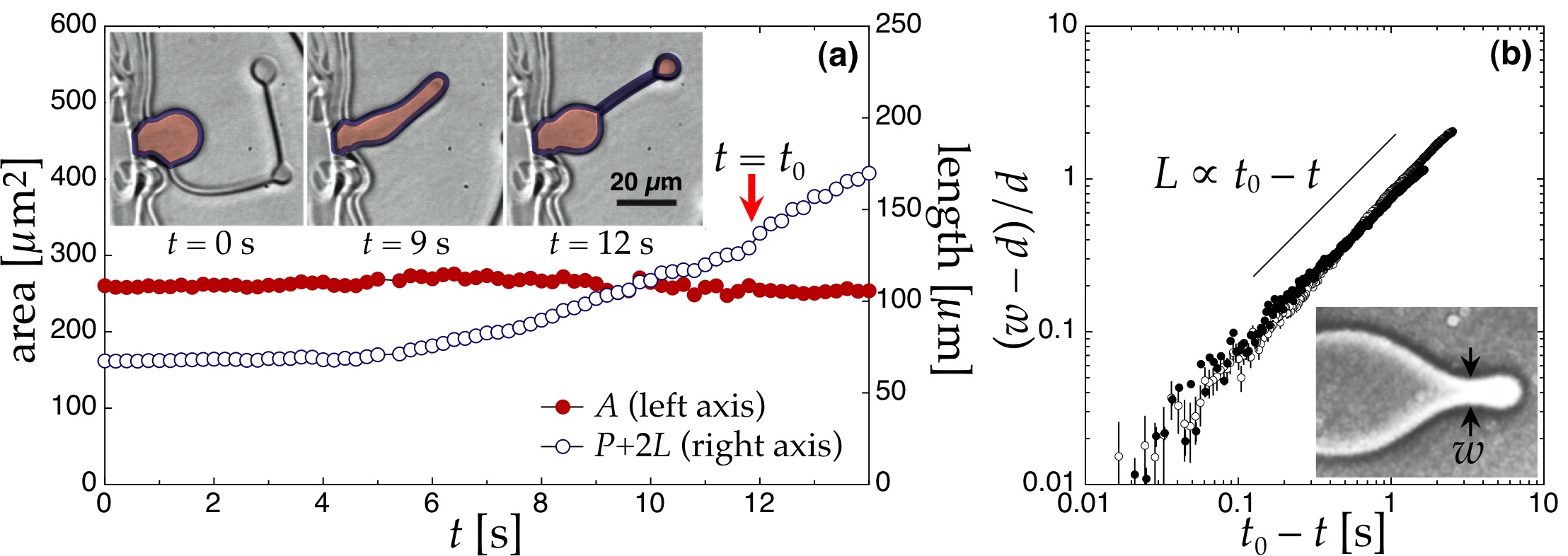}
\caption{(color online) (a) The deformation of a pancake to form a myelin. $A$ is the pancake area (shaded); $P$, $L$ are respectively the pancake perimeter and the myelin length (traced in thick outline); the myelin forms at $t=t_{0}$. (b) The width $w$ at the waist of the elongated pancake as a function of $t_{0}-t$.}
\label{pancakeAP}
\end{figure*}	

\section{Myelin instability: flow and buckling}

The drying drop experiment provides the clearest picture of how a myelin is formed: the initial structure is well-ordered, the dynamics are slow, and the view is unobstructed. In this experiment, isolated, slowly growing myelins are formed from small, planar surfactant ``pancakes'' situated at the pinned contact line of a slowly evaporating water drop. The process follows a characteristic sequence: the initially rounded pancake becomes elongated, develops a pronounced ``waist'' which quickly narrows down to a cylinder --- the myelin (Fig.~\ref{pancakeAP}a, insert). The entire event take place in $\sim 10$ s, at the end of which the original pancake has transformed into a pancake with a myelin growing from its leading edge. 	

By tracing the pancake/myelin outline with a multipoint spline, the pancake morphology can be easily monitored as the myelin forms. Although the formation of a myelin represents a drastic change in morphology, the area $A$ of the pancake remains constant throughout. Meanwhile the combined boundary length $P+2L$, where $P$ is the perimeter of the pancake alone and $L$ the myelin's length, smoothly increases across the transition (Fig.~\ref{pancakeAP}a). This suggests myelin formation may be thought as the deformation of a 2D shape (the pancake) of fixed area subjected to a continuously increasing boundary, with the myelin being made up of two lengths of excess perimeter. 

There is a surprising resemblance between myelin formation and the break-up of a pendant liquid drop. Both start with a rounded shape which elongates and develops a rapidly narrowing waist --- although the drop eventually breaks-up while the myelin stays attached. In viscous drop break-up, the radius $r$ of the waist narrows in time as $r\propto t_{0}-t$ \cite{Cohen&c1999}, where $t_{0}$ is the instant of break-up. In myelin formation, using high speed (250 images/s) video microscopy, the waist $w(t)$ of the pancake is found to narrow in a similar way,
\begin{equation}
w-d \propto (t_{0}-t)^{1.1\pm 0.1} \ ,
\end{equation}
(Fig.~\ref{pancakeAP}b); here $d$ is the myelin diameter, and $t_0$ is the instant when the myelin formed.

The resemblance between myelin formation and drop break-up, I suggest, is not coincidental but reflect similar underlying physics. The pendant drop is a 3D object of fixed volume. Gravity stretches the drop, surface tension resists the deformation, and the competing forces lead to a Rayleigh instability, breaking up the drop. The pancake is a quasi-2D object of fixed area. It is stretched by surfactant flow, bilayer elasticity resists this deformation, and the competing forces lead to mechanical instability, forming a myelin. As I hope to show, in myelin formation, as in drop break-up, the competition between a deforming body force and a restorative boundary force leads to an instability and rapid morphological change.

Here, the mixed mechanical properties of the surfactant lamellar structure are crucial. A bilayer is a fluid membrane within which surfactant molecules can freely diffuse; but the bending the bilayer costs elastic energy 
\begin{equation}\label{bending_energy}
\mathcal{F}^\mathrm{B} = \frac{\kappa}{2}\int\!\!\!\!\int \left(C_1+C_2\right)^2 dS \ ;
\end{equation}
here $\kappa$ ($\sim 20 \ k_\mathrm{B}T$) is the bending modulus, $C_1$, $C_2$ the principle curvatures, and the integral is over the entire membrane. Stacked into a lamellar structure, the bilayers maintain a preferred repeat spacing, resisting compression and dilation; moreover, for poorly-solubale myelin-forming surfactants, material exchange between disconnected bilayers is negligible \cite{Israelachvili1992}. Thus laterally, parallel to the bilayers, the lamellar structure acts as a fluid; but in the transverse direction it is an elastic solid.

\begin{figure*}[tb]
\center
\includegraphics[height=2.2in]{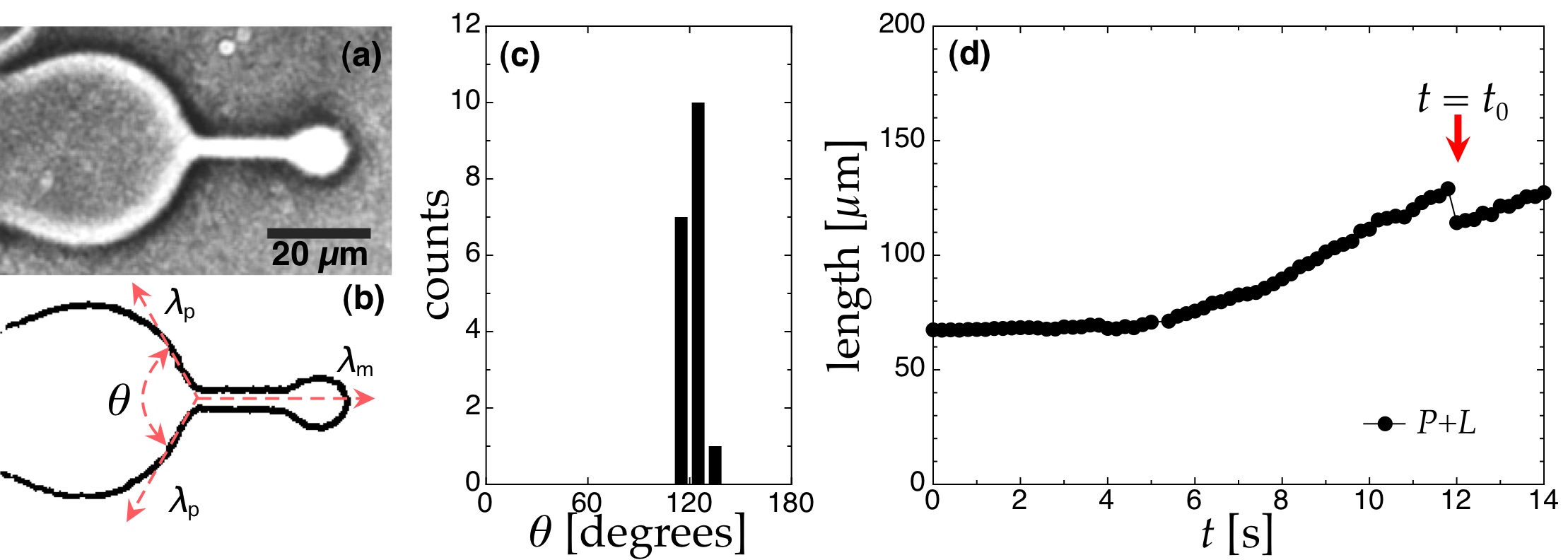}
\caption{(color online) (a) The pancake/myelin junction. (b) Local balance of forces determines the static junction angle $\theta$. (c) The distribution of $\theta$ observed in experiment. (d) The combination $P+L$ as a function of time; $P$ is the pancake perimeter, $L$ is the myelin length.  $P+L$ is clearly discontinuous at $t_{0}$, when the myelin forms. This is in contrast to Fig.~\ref{pancakeAP}a, which shows $P+2L$ is continuous across $t_0$.}
\label{angles}
\end{figure*} 

A pancake astride the pinned contact line of a water drop is more hydrated at its leading edge, inside the drop, than at its rear. As described in Section III, we can describe myelin growth on the basis of surfactant flow driven by the hydration gradient across the pancake; the same flow of surfactant from the rear of the pancake towards its leading edge can also stretch and deform the pancake, leading to myelin formation. Additionally, such a flow also increases the overall hydration of the pancake. As Huang \textit{et al.} showed, even a slight increase in overall hydration can overcome the added elastic energy of forming a myelin \cite{HuangZouWitten2005}. In this sense, the hydration gradient plays the role in myelin formation analogous to the force of gravity in pendant drop break-up.

\begin{figure*}[t]
\center
\includegraphics[height=2.2in]{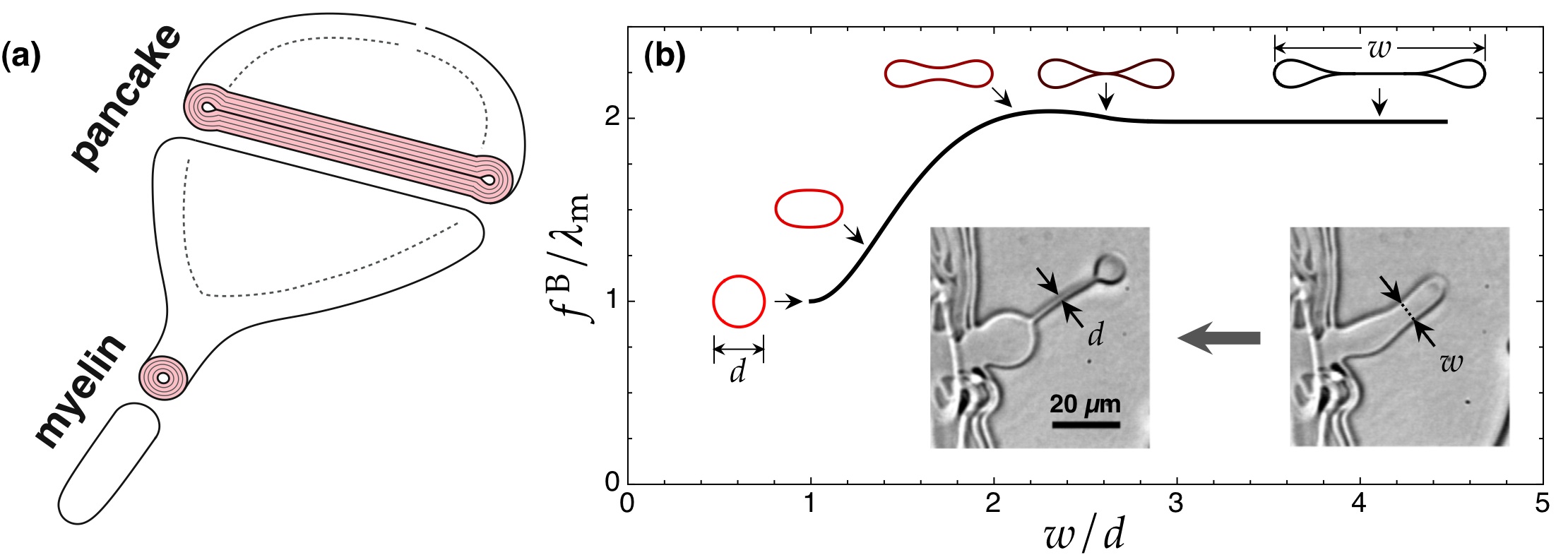}
\caption{(color online) (a) Schematic drawing of a pancake/myelin complex, with transverse cross sections (shaded) showing its nested lamellar structure. The transverse cross section of the pancake looks like that of a collapsed elastic tube, while the transverse cross section of the myelin is circular. (b) The cross-sectional bending energy $f^\mathrm{B}$ as a function $w$, the width of the transverse cross-section; $d$ is the diameter of the resulting myelin. Shown above the curve is a sequence of calculated cross sections that would be obtained by a transverse cut across the narrow waist of the elongated pancake, as indicated by the arrows in the photographic insert. These show the transition from elongated pancake (wide cross section that is collapsed in the middle) to the myelin (circular cross section); the transverse cross section narrows in width approaching myelin formation, but the area $a$ it encloses is constant. }
\label{structure}
\end{figure*}

Confocal microscopy indicate the pancake is composed of multiple nested, closed bilayers enclosing little free volume \cite{ZouNagel2006}.  Therefore even a flat pancake has some bending energy because the bilayers are sharply folded at the pancake's edge (Fig.~\ref{structure}a). Since the bilayer curvatures are localized at the pancake's edge, and since the pancakes are thin ($1-2 \ \mu$m), a pancake's total bending energy is just proportional to its perimeter, $\mathcal{F}^{{B}}_\mathrm{p}=\lambda_\mathrm{p} P$; $\lambda_\mathrm{p}$ is then the pancake's line tension, the 2D analog to the surface tension of a liquid drop. We can see its effect in the pancakes' rounded shapes prior to myelin formation. Presumably, $\lambda_\mathrm{p}$ also acts to resist the deformation of the pancake caused by surfactant flow. 	 

To complete the analogy with pendant drop break-up, what instability, if any, does the competition between extensional surfactant flow and pancake line tension lead to? There is no Rayleigh instability in 2D fluid flow, however, the geometry of the myelin/pancake junction (Fig.~\ref{angles}a) hints at the existence of a mechanical instability.  The pancake has a line tension $\lambda_\mathrm{p}$; similarly, the myelin has a line tension (bending energy per unit length) $\lambda_\mathrm{m}$. Local force balance at the junction between pancake and myelin leads to the relation
\begin{equation}\label{junction_angle}
\lambda_\mathrm{m} = 2\lambda_\mathrm{p} \cos(\theta/2)\ ,
\end{equation}
where $\theta$ is the static junction angle (Fig.~\ref{angles}b). Eq.~\ref{junction_angle} has a solution for $\theta$ only if $\lambda_\mathrm{m}\leq 2\lambda_\mathrm{p}$, \textit{i.e.} forming a myelin from two lengths of perimeter leads to a reduction in bending elastic energy.  Experimentally, $\theta$ is measured to be $110^{\circ}$-$130^{\circ}$ (Fig.~\ref{angles}c), which implies $\lambda_\mathrm{m}\approx\lambda_\mathrm{p}$. If this implication is true, then the total elastic energy of the pancake plus myelin is proportional to the pancake perimeter plus myelin length,
\begin{equation}
\mathcal{F}^\mathrm{B}_{\mathrm{p+m}} = \lambda_\mathrm{p} P + \lambda_\mathrm{m} L \propto P+L \ ,\ \textrm{if} \ \lambda_\mathrm{p} \approx \lambda_\mathrm{m} \ .
\end{equation}
As a function of time, $P+L$ is discontinuous at $t_{0}$, the instant of myelin formation: it drops sharply, before resuming growth at a slower rate (Fig.~\ref{angles}d). This suggests $\mathcal{F}^\mathrm{B}_{\mathrm{p+m}}$ is discontinuous at $t_{0}$, indicative of an instability.

The instability identified here is related to the collapse of an elastic tube. The transverse cross section of an elongated pancake, as sketched in Fig.~\ref{structure}a, should be the same as that of a collapsed elastic tube (\textit{e.g.} a soft water hose when the pressure head is turned off). This is because transverse cross sections of (thin-walled) elastic tubes are closed curves with minimal bending energies \cite{TadjbakhshOdeh1967}. Starting with a unperturbed tube with a circular cross section, by lowering the pressure $p$ inside the tube, the tube will eventually collapse: the cross section will have a flat region of wall-to-wall contact in the middle, and teardrop-shaped bulges at the edges \cite{FlahertyKellerRubinow1972, KreschNoordergraaf1972}. The shape of the transverse cross-sectional is controlled not by internal pressure \textit{per se} but the ratio $\varepsilon \equiv c^2/a$, where $c$ is the cross-sectional circumference, and $a$ the cross-sectional area.  For an elastic tube, $c$ is fixed; what lowering $p$ does is to reduce $a$. This is how the tube cross section evolves from a circle, where $\varepsilon$ is minimal, to a collapsed shape, where $\varepsilon$ is large. 

In myelin formation, the constraints are reversed. Here, because the pancake has a fixed thickness (lamellar structure is a solid in the transverse direction), the area $a$ enclosed by a transverse cross section must be fixed. Instead, because a pancake is laterally fluid and can flow, as it elongates, the width $w$ --- and circumference $c$ --- of a transverse cross-section in waist area will decrease. For a sufficiently elongated pancake, its width at the waist can become too narrow ($\varepsilon$ too small) to sustain a collapsed shape, and the circular transverse cross section of a myelin will become favored.

In the linear elastic regime, the family of transverse cross sections from pancake to myelin (Fig.~\ref{structure}b), where $c$ is variable but $a$ is fixed, can be obtained numerically; I follow the calculation of Flaherty \textit{et al.} for collapsing elastic tubes \cite{FlahertyKellerRubinow1972}, modified to reflect the constraints appropriate to myelin formation (see Appendix). The bending energy of each transverse cross section
\begin{equation}\label{cross_sectional_energy}
f^\mathrm{B} =\frac{\kappa}{2}\oint C^2 ds \ 
\end{equation}
(here $C$ is the curvature, and the integral is along the cross-sectional circumference) can then be evaluated as a function of the cross-sectional width $w$. Plotted against $w$, $f^\mathrm{B}$ flattens to a constant ($=2\lambda_\mathrm{p}$) when $w$ is large; but when $w \lesssim 2.3d$, the collapsed cross section is unstable: $f^\mathrm{B}$ falls with decreasing $w$, reaching its minimum value ($=\lambda_{\mathrm{m}}$) at $w=d$ and a circular cross section --- a myelin (Fig.~\ref{structure}b). This calculation indicates $\lambda_{\mathrm{m}}\approx\lambda_{\mathrm{p}}$, consistent with the estimate made on the basis of balance of forces at the myelin/pancake junction.  Inserting the calculated value of $\lambda_{\mathrm{m}}/\lambda_{\mathrm{p}}$ into Eq.~\ref{junction_angle} yields a myelin/pancake junction angle $\theta = 119^\circ$, in good agreement with observation.

If the instability described above is indeed responsible for myelin formation, then its dynamics should be controlled by surfactant flow and bilayer elasticity. Just as dimensional analysis can be used to analyze the dynamics of fluid instabilities, e.g. drop break-up, a similar analysis can be applied to myelin formation. The relevant physical parameters here are the 2D pancake viscosity $\eta_\mathrm{p}$ (dimension MT$^{-1}$), the line tension $\lambda_\mathrm{p}$ (dimension MLT$^{-2}$), the reduced waist width $w - d$, and $t_{0}-t$. From these one can construct only a single dimensionless group,
\begin{equation}
\frac{\lambda_\mathrm{p}}{\eta_\mathrm{p}}\left(\frac{t_{0}-t}{w - d}\right) \quad \Rightarrow \quad 
w - d \propto t_{0}-t\ .
\end{equation}
The implied linear dependence of $w$ on $t_{0}-t$ is consistent with experimental observations. One may further estimate the velocity $\lambda_\mathrm{p}/\eta_\mathrm{p}$. Taking $\lambda_\mathrm{p} = \lambda_\mathrm{m}\sim N\kappa/d$, where $N\approx 100$ is the number of nested bilayers, $\kappa \approx 10^{-19} \ \textrm{J}$ is the bending modulus of a single bilayer, and $d \approx 1 \ \mu\textrm{m}$ is the myelin diameter; and taking $\eta_\mathrm{p} \sim N\eta$, where $\eta \approx 10 \ \textrm{cP}\cdot\mu\textrm{m}$ is the 2D viscosity of a single bilayer \cite{EvansNeedham1987}, one obtains $\lambda_\mathrm{p}/\eta_\mathrm{p} \sim 10 \ \mu\textrm{m/s}$. This compares well to the experimentally measured velocity ($\approx 1 \ \mu\mathrm{m/s}$).

Finally, this model suggests the hydration gradient is not the only force which can lead to myelins. Any body force (such as shear stress) can also lead to myelin formation, as long as it can sufficiently stretch the pancake. Indeed, myelins can be induced to form by shear flows even in the absence of hydration gradients \cite{ZouNagel2006}.

\section{Summary}

In this paper, I argue the fluid flow of surfactant, driven by the hydration gradient, results in the formation and growth of myelins. The immediate consequence of this argument is there can be no myelins without surfactant flow. This give a natural explanation to the well-known observation that myelins do not form for temperatures below the so-called main transition temperature $T_\mathrm{m}$ ($0^\circ \textrm{C}$ for DLPC and $23^\circ\textrm{C}$ for DMPC) \cite{ChapmanFluck1965}. Below $T_\mathrm{m}$, the surfactant tails freeze into a solid phase and the lamellar structure loses its lateral fluidity; consequently myelins can no longer form.

The varying growth behaviors as seen in the different experiments, I suggest, is entirely due to the different ways the driving force, the hydration gradient, relaxes in a given situation. This is a function of geometry, as well as initial and boundary conditions. A good test of the model suggested in Section III is to devise additional experimental geometries, and to examine myelin growth in those instances.

While the fluid flow of surfactant is crucial, I suggest the initial formation of myelin figures also depends on the elastic character of the surfactant lamellar structure. In analogy to a classic fluid instability --- the break-up of a pendant drop --- myelin formations is a result of the competition between bulk surfactant flow, driven by the hydration gradient, and bilayer bending, manifesting itself as a line tension. The instability at its heart is closely related to a classic elastic instability --- the buckling and collapse of an elastic tube (in reverse). The elastic forces involved here may be measured directly. By pulling on a well-ordered, equilibrium surfactant lamellar pancake until a myelin is formed, using possibly an optical tweezer or a microcantilever, one should be able to determine the pancake edge tension $\lambda_\mathrm{p}$ and the myelin line tension $\lambda_\mathrm{m}$ from the work of deformation. This would be a direct test of the myelin instability mechanism suggested here.
 
The experiments and calculations described here aim to show that the formation of myelin depends crucially on the mixed mechanical properties of the surfactant lamellar structure: the wet soap has to be able to flow as a fluid \textit{and} buckle like an elastic solid. Mechanical instabilities in fluids (\textit{e.g.} the Rayleigh instability) and elastic solids (\textit{e.g.} buckling) have been active subjects of study since the work of Euler; myelin formation is of a rather less familiar type: a mechanical instability of mixed character in a material where fluid and elastic properties coexist. A large number of similarly mixed instabilities must surely exist, given the diversity of liquid crystalline phases, both thermotropic (temperature controlled) and lyotropic (as in soap). In particular, some of the diverse interfacial structures seen in surfactant dissolution may be understood in terms of mechanical instabilities of the interface.

Finally, I have not treated an especially striking instability of the myelins themselves: coiling and the formation of helices (see Fig.\ref{Myelin_Dark_Field}). The coiling often appears spontaneously, and can also be induced by the addition of polymers or $\textrm{Ca}^{2+}$ ions \cite{Haran&c2002, SakuraiSuzukiSakurai1990, Frette&c1999, LinWeisMcConell1982}. The case of polymer-induced coiling has been attributed to the appearance of a spontaneous membrane curvature in the presence of the additive. But in general one expects that myelins, being laterally fluid, cannot support the torsion needed to form helices and coils. The treatment of myelins in this paper is simplistic in the sense I assume the myelin is perfectly fluid laterally and perfectly elastic in the transverse direction. While such a simple picture appears sufficient when it comes myelin formation and growth, the coiling instabilities requires more elaborated treatments \cite{SantangeloPincus2002, Huang2006}.

\appendix 

\section{Calculating the Elastic Energies}

\subsection{Linear elasticity: thin-walled elastic tube}

The calculation for the transverse cross section of pancakes and myelins is based on the calculation of Flaherty \textit{et al.} \cite{FlahertyKellerRubinow1972} for collapsed elastic tubes; it thus useful to briefly outline the physics which gives us the cross sections of elastic tubes.

The calculation of Flaherty \textit{et al.} assumes linear elasticity. For a tube made from an conventional elastic material, linear elasticity holds if strains are small; practically, this is satisfied if the tube wall is thin. Then, the equations of mechanical equilibrium, which one solve to find the tube cross sections $\mathbf{x}_\mathrm{t}(s)$, can be derived from the theory of thin elastic shells \cite{KreschNoordergraaf1972}. They can also be derived from minimizing the energy functional 
\begin{equation}\label{thin_tube}
\mathcal{H}[\mathbf{x}_\mathrm{t}(s)] = \frac{\kappa}{2}\oint \left(C[\mathbf{x}_\mathrm{t}] - 1/r_\mathrm{t}\right)^{2} ds - \frac{p}{2} \oint \hat{n}\cdot \mathbf{x}_\mathrm{t} \ ds \ ;
\end{equation}
here $C$ is the radius of curvature, $r_\mathrm{t}$ is the radius of the unperturbed tube, $p$ is the pressure difference between the interior and exterior of the tube, and $\hat{n}$ in the outward unit normal \cite{TadjbakhshOdeh1967}. One seek solutions $\mathbf{x}_\mathrm{t}(s)$ of the form
\begin{equation}
\mathbf{x}_\mathrm{t}(s)=\int_{0}^{c}\left[\begin{array}{r}
       \cos\Theta(s)  \\
      \sin\Theta(s)   
\end{array}\right] ds \ , \quad C(s)\equiv\Theta'(s) \ ; 
\end{equation}
here $c=2\pi r_\mathrm{t}$ is the tube's cross-sectional circumference. Since $c$ is fixed for an elastic tube, one further requires
\begin{equation}
\mathbf{x}_\mathrm{t}(s)=\mathbf{x}_\mathrm{t}(s+c), \quad 
\Theta(s+c) = \Theta+2\pi \ .
\end{equation}

The second term in $\mathcal{H}$, involving $p$, is simply $-pa$, where $a$ is the area enclosed by $\mathbf{x}_{\mathrm{t}}$. This term constrains the tube cross section, whose circumference $c$ is fixed, to enclose an area $a$ (which is set by the pressure difference $p$). The first term in $\mathcal{H}$, involving the curvature $C$, is the elastic energy of the transverse cross section; this term is just $f^{B}$, the cross-sectional bending energy as defined in Eq.~\ref{cross_sectional_energy}, plus a constant. Thus $\mathbf{x}_{\mathrm{t}}$ is a cross section of circumference $c$ (which is fixed), enclosing an area $a$ (which varies as a function of $p$), whose cross-sectional bending energy $f^{B}$ is minimal. Finally $p$, being a dimensional quantity, cannot be the parameter which characterizes the cross sectional \textit{shape}; instead, that role is played by the dimensionless ratio $\varepsilon=c^2/a$.

The transverse cross sections of the pancake, like those of the elastic tube, minimizes the cross-sectional bending energy for a given $\varepsilon$. But here the constraint to shape evolution is different: $a$ is fixed and $c$ varies as cross-sectional width $w$ varies. The control parameter is still $\varepsilon$, and so we can obtain a pancake cross section simply by re-scaling a similarly shaped elastic tube cross section. If $\mathbf{x}_\mathrm{t}(s)$ is an elastic tube cross section with fixed circumference $c$ enclosing a variable area $a$, a pancake cross section $\mathbf{x}_\mathrm{p}(s')$ of the same shape (having the same value of $\varepsilon$), with variable circumference $c'$ enclosing a fixed area $a'$, is given by
\begin{equation}\label{scaling}
\mathbf{x}_\mathrm{p}(s') = \left(\frac{a'}{a}\right)^{1/2} \mathbf{x}_\mathrm{t}(s) \ .
\end{equation}
Once family of elastic tube cross sections has been found, applying Eq.~\ref{scaling} will yield a corresponding family of pancake cross section, whose widths and cross-sectional bending energies can be evaluated. The curve in Fig.~4b was calculated from 1500 pancake cross sections, whose the cross sectional width $w$ was varied systematically from $1\times$ to $\sim4\times$ the myelin diameter $d$.

\subsection{Thick walls: effective bending modulus $\kappa_\mathrm{eff}$}

\begin{figure}[tbh]
\center
\includegraphics[width=2.8in]{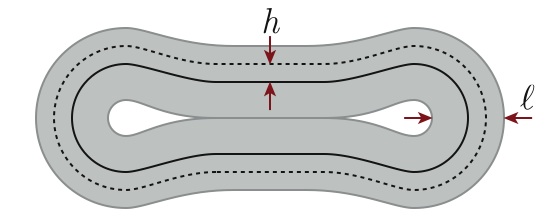}
\caption{(color online) The transverse cross section of a pancake with wall-thickness $\ell$. The solid line is the median surface $\bar{s}$; the dashed line $s(h)$ is a surface (bilayer) displaced by $h$ from $\bar{s}$.}
\label{thickwall}
\end{figure}

The agreement between the previous calculation, based on thin-walled tubes, and experimental observation of myelin formation may be surprising. With a thin-walled tube, any radius of curvature will be large compared to wall thickness, therefore strains ares small and linear elasticity applies. With thick-walled tubes there can be large strains and nonlinearities.

For a conventional elastic material, if a slab of thickness $\ell$ is bent through a radius of curvature $R$, the maximum strain within the slab will be on the order of $\ell/R$; when $R\sim \ell$, the maximum strain is of order 1 and the linear elasticity breaks down. This means the slab's bending modulus will be function of the radius of curvature rather than a constant.

It may appear that the previous calculation cannot apply to a thick-walled pancake with $N\sim 100$ bilayers. However, a lamellar structure of surfactant bilayers is not a conventional elastic material: the bilayers are fluid, and adjacent bilayers are free to slide past each other. Thus the bending of each bilayer can be considered independently, subject to only the requirement of a fixed bilayer repeat spacing $d_{0}$; the total cross-sectional bending energy of a $N$-layer pancake is just the sum of each bilayer's individual contribution. We can rewrite this sum in terms of the bending of a single median surface $\bar{s}$ and a curvature-dependent effective bending modulus $\kappa_\mathrm{eff}$
\begin{equation}
f^\mathrm{B} =\frac{\kappa}{2}  \sum_{i=1}^{N} \oint C_{i}^{2} ds_{i}\equiv \frac{1}{2} \oint \kappa_\mathrm{eff}\bar{C}^{2} d\bar{s} \ .
\end{equation}
For large $N$, we can make the continuum approximation where the bilayer $s_{i}\rightarrow s(h)$ is labelled by its displacement $h$ from the median surface,
\begin{equation}\label{total_energy_thick_cont}
f^\mathrm{B} = \frac{\kappa}{2} \int_{-\ell/2}^{\ell/2} \oint \frac{ds(h)}{\left(\bar{R} + h\right)^{2}}\frac{dh}{d_{0}} \equiv \frac{1}{2} \oint \frac{\kappa_\mathrm{eff}}{\bar{R}^{2}}d\bar{s} \ ;
\end{equation}
here $\ell=Nd_{0}$ and $\bar{R} = 1/\bar{C}$ is the radius of curvature at the median surface (Fig.~\ref{thickwall}).

\begin{figure}[tbh]
\center
\includegraphics[width=2.8in]{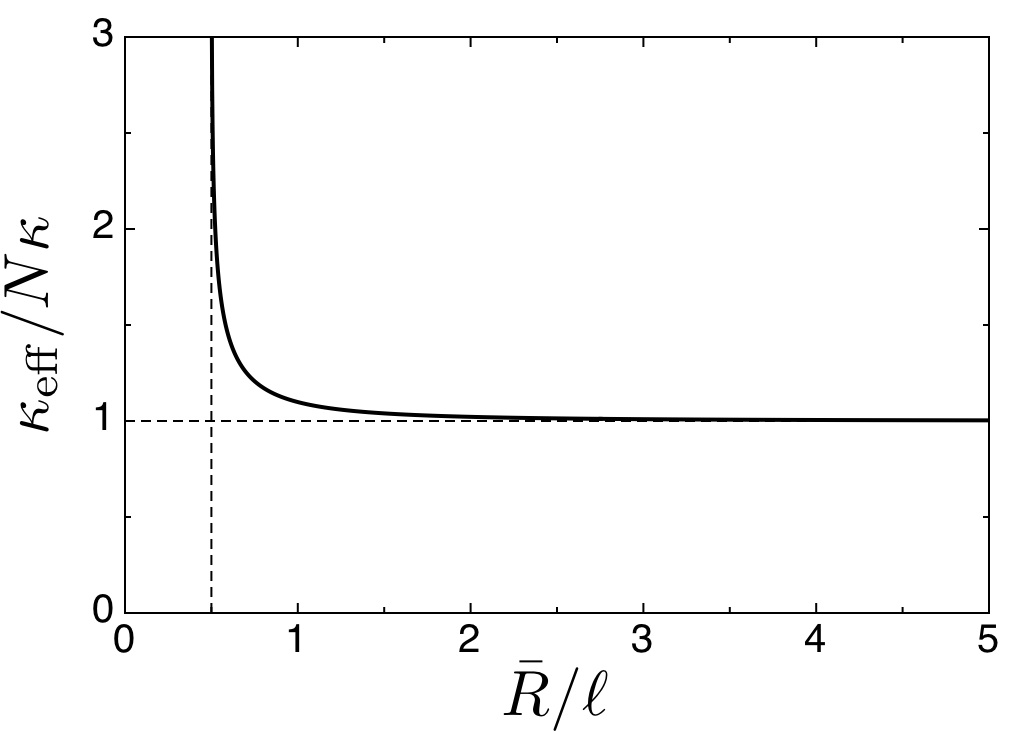}
\caption{The effective bending modulus $\kappa_\mathrm{eff}$ of $N$ bilayers as a function of $\bar{R}$, the median radius of curvature; $\ell=Nd_{0}$ is the total thickness of the $N$-bilayer lamellar structure.}
\label{keff_plot}
\end{figure}

If the curvature $\bar{C}$ on $\bar{s}$ is approximately constant over length scales comparable to $\ell$, then due to the fact the lamellar repeat spacing is fixed, the differential arc lengths $ds(h)$ on surface $s(h)$ and $d\bar{s}$ on $\bar{s}$ are related by 
\begin{equation}
ds(h) = \frac{\bar{R}+h}{\bar{R}} d\bar{s} \ .
\end{equation}
Inserting this relation into Eq.~\ref{total_energy_thick_cont} yields the following expression for $\kappa_\mathrm{eff}$,
\begin{equation}
\kappa_\mathrm{eff} =  \kappa \left(\frac{\bar{R}}{d_{0}}\right)\log\left(\frac{\bar{R} +\ell/2}{\bar{R}-\ell/2}\right)
\end{equation}
As shown in Fig.~\ref{keff_plot}, $\kappa_\mathrm{eff}$ has significant variations with $\bar{R}$ only close to the singular limit of $\bar{R}=\ell/2$. Therefore a constant bending modulus independent of local curvature, which is a consequence of linear elasticity, should remain an excellent approximation. The analysis for myelin formation based on thin-walled tubes should hold even for thick-walled pancakes with many bilayers.

\begin{acknowledgments}
I am grateful to Sidney Nagel for his advice and support, and to Jung-Ren Huang and Tom Witten for teaching me about soap. This work was funded by NSF MRSEC DMR-0820054, NSF DMR-0652269, and the Keck Foundation.
\end{acknowledgments}

\end{document}